\documentclass{jssc}

\def\textsubscript#1%
{$_{\text{#1}}$}

\def\cdd{\mbox{\boldmath$\cdot$}~}

\def\ay{\arraycolsep=1.5pt}

\input{amssym.def}
\include{graphix}
\usepackage{titlesec}
\usepackage{enumerate}
\usepackage{amssymb}
\usepackage{graphicx} 
\usepackage{epstopdf}
\usepackage{caption}
\usepackage{algorithm}
\usepackage{algorithmic}
\usepackage{amsmath}

\usepackage{array}

\newcommand{\MC}{\mathcal{C}}

\newcommand{\bF}{\mathbb{F}}

\makeatletter
\def\@oddfoot{\hfill}
\newcount\shumeicount
\def\setshumei#1#2#3{%
  \shumeicount=\count0
  \def\@oddhead{%
    \raise-5pt\hbox to0pt{\vrule width\hsize height 0pt depth 0.4pt\hss}\relax
    \ifnum \shumeicount=\count0
      \raise-7pt\hbox to0pt{\vrule width\hsize height 0pt depth 0.4pt\hss}\relax
      #1
    \else
      \ifodd\count0
        #2
      \else
        #3
       \fi
     \fi
  }%
}
\makeatother
\makeatletter
\def\@oddfoot{\hfill}
\newcount\shujiaocount
\def\setshujiao{%
  \shujiaocount=\count0
  \def\@oddfoot{%
      \ifodd\count0
      \else
      \fi
  }%
}
\makeatother
\def\title#1#2#3#4{{
  \vspace*{0.3cm}
  \begin{flushleft} \Large\bf #1\end{flushleft}
  \vspace*{-0.2cm}
      \begin{flushleft}
      \bf #2
      \end{flushleft}
      \footnotetext{\hspace{-6mm} #3\\ #4}}}

\def\dshm#1#2#3#4#5
{\setshumei{J Syst Sci Complex #1, #2(#3):
{\thepage--\pageref{LastPage}}\hfill}
            {\hfill {\small #4}\hfill\hbox to0pt{\hss\thepage}}
            {\hbox to0pt{\thepage\hss}\hfill {\small #5}\hfill
            }
            \setshujiao}
\def\drd#1#2
{{\vskip 1cm\small \begin{flushleft}
 #1 \\
 #2\\
\copyright The Editorial Office of JSSC \&  Springer-Verlag GmbH
Germany 2024
\end{flushleft}}}

\def\epsilon{\varepsilon}

\begin{document}

\title{Linear Complementary Dual Codes Constructed from Reinforcement Learning$^*$}
{\uppercase{Wu} Yansheng \cdd \uppercase{Ma} Jin \cdd \uppercase{Yang} Shangdong}
{\uppercase{Wu} Yansheng \\
School of Computer Science, Nanjing University of Posts and Telecommunications, Nanjing, $210023$, China;
Email:yanshengwu@njupt.edu.cn\\
\uppercase{Ma} Jin   \cdd \uppercase{Yang} Shangdong \\
School of Computer Science, Nanjing University of Posts and Telecommunications, Nanjing, $210023$, China.  \\
Emai: 1222046136@njupt.edu.cn; sdyang@njupt.edu.cn.}
{$^*$\rm This work was supported by the National Natural Science Foundation of China (Nos. 62372247, 12101326, 62206133)  and  the Open Project of Guangxi Provincial Key Laboratory (No. MIMS22-01).\\
{$^\diamond${\it This paper was recommended for publication by Editor . }}
 }

\drd{DOI: }{Received: 04 Jul 2024}{ / Revised: 03 Sep 2024}


\dshm{2024}{XX}{ }{CONSTRUCT LCD CODE BASED RL}{\uppercase{Wu} Yansheng 
$\cdd$ \uppercase{Ma} Jin $\cdd$ \uppercase{Yang} Shangdong}

\Abstract{Recently, Linear Complementary Dual (LCD) codes have garnered substantial interest within coding theory research due to their diverse applications and favorable attributes. This paper directs its attention to the construction of binary and ternary LCD codes leveraging curiosity-driven reinforcement learning (RL). By establishing reward and devising well-reasoned mappings from actions to states, it aims to facilitate the successful synthesis of  binary or ternary LCD codes. Experimental results indicate that LCD codes constructed using RL exhibit slightly superior error-correction performance compared to those conventionally constructed LCD codes and those developed via standard RL methodologies. The paper introduces novel binary and ternary LCD codes with enhanced minimum distance bounds. Finally, it showcases how Random Network Distillation aids agents in exploring beyond local optima, enhancing the overall performance of the models without compromising convergence.}      




\section{Introduction}

The rapid development of data communications has increasingly underscored the necessity for reliable data transmission. In this context, Claude Shannon published his groundbreaking work \lq\lq A Mathematical Theory of Communication\rq\rq \cite{1}, which laid the foundational groundwork for modern information theory and coding theory. Shannon posited that every communication channel possesses a parameter known as channel capacity; if the transmission rate demanded by a communication system is less than this capacity, there exists a coding scheme such that when applied with an appropriate code length and maximum likelihood decoding, the probability of error in the system can approach zero arbitrarily.

However, while Shannon established the existence theorem for such codes, he did not provide explicit instructions on how to construct them. This challenge spurred further research in the field of information theory, particularly in the design of error-correcting codes. Researchers, inspired by Shannon's theorem, diligently sought innovative coding strategies to enhance the reliability of data transmission.

Over the subsequent decades, scholars have devised numerous classes of error-correcting codes with strong performance metrics. Yet, these high-performance codes typically embody specific structures or adhere to certain properties. Without such structures, the utility or error-correcting capability of a code may be significantly limited.

Linear Complementary Dual (LCD) codes, as a special class of linear codes, have garnered substantial interest due to their unique property that their intersection with their dual codes contains only the zero element. This inherent characteristic makes LCD codes a focal point of ongoing research, as they offer potential advantages in terms of resistance to errors and security enhancements in communication systems.

Currently, the construction of LCD codes predominantly relies on principles within coding theory, where the design challenge is reframed as an optimization problem of code parameters to control the code's performance. For LCD codes, three essential parameters are typically considered: the code length $n$, dimension $k$, and minimum distance $d$. If $\MC$ is a nonempty subset of an $n$-dimensional linear space over $\bF_q$, $\MC$ is said to be a $q$-ary error-correcting code. Each element in the code $\MC$ is called a codeword of $\MC$. Let $k$ represent the information bits of the code, and $k/n$ represent the code rate, which is the basic parameter to measure the effectiveness of the code.
The minimum distance $d$ refers to the smallest Hamming distance between any two distinct codewords, indicating the code's error-correcting prowess. These three parameters interact. For example, under a fixed code length $n$, increasing the dimension $k$ might result in a smaller minimum distance. Their mutual constraints can be mathematically represented through inequalities, thereby establishing bounds for the parameters. Codes that closely approach or reach these bounds are generally considered to have superior performance.

Based on coding theory, different codes and construction methods require specific expertise, and without considering decoders or channel conditions during construction, accurate assessment of code performance is not possible. Moreover, certain coding theory approaches involve constraint conditions; for instance, constructing truncated LCD codes may necessitate the use of distinct theorems depending on whether the codeword weights are odd or even, or even the parity of the dimension, which significantly impacts the flexibility in constructing LCD codes \cite{2}. Finally, construction methods rooted in coding theory often concentrate exclusively on a singular performance metric, thereby encountering challenges in simultaneously optimizing multiple criteria, rendering comprehensive consideration intricate. And many methods for computer implementation are more difficult, can not use computer resources.

With the development of artificial intelligence technology, the combination of error-correcting code and artificial intelligence has become an effective method to build error-correcting code. This kind of method can not only automate coding, access channel conditions and decoder modules, accurately evaluate the performance of the code, but also access various modules, such as curiosity module, with high scalability. Kim et al. proposed ``deepcode" based on AWGN feedback channel and conducted unified training for encoders and decoders based on recurrent neural networks \cite{3}.They demonstrate that the conjunction of deep learning methodologies with channel-specific insights has the potential to excel over coding systems meticulously refined through decades of rigorous mathematical inquiry. Gruber \cite{4} also found that structured codes are easier to learn, and neural networks can generalize structured codes, which means that structured codes have more advantages in learning. Recently, some scholars have also applied artificial intelligence techniques to coding theory, which is called AI coding \cite{5}. However, the code constructed in \cite{5} is limited to ordinary linear code, does not contain some structure, and only focuses on the case of binary field, which makes its usability limited.

Based on the properties of LCD codes, this paper presents a method of constructing binary and ternary LCD codes based on reinforcement learning (RL) from the perspective of reward function and action mapping. This makes it possible for RL to construct any kind of code, and is no longer limited to the binary field, further improves the usability of RL in error-correcting coding, and accelerates the cross-application of RL and error-correcting coding. RL shares remarkable similarities with the challenges encountered in constructing error-correcting codes, particularly in confronting the necessity to navigate through immense state spaces. Employing the principles of RL, in order to deal with the challenge of excessive code space, the random network distillation (RND) algorithm \cite{6} is added to the evaluation module to generate internal incentives, and curiosity is added to the agent to assist exploration. By using the method proposed in this paper, the existence of some  almost optimal LCD codes is found, and the range of the minimum distance bound of some LCD codes is reduced.

\subsection{Related work: LCD codes}

LCD codes, first proposed by Massey \cite{7} in 1992, have been the focus of extensive research because of their unique theoretical properties and practical significance. From a theoretical point of view, Massey proved that LCD codes can provide an optimal encoding of a two-user binary addition channel \cite{7}. Sendrier \cite{8} proved that LCD codes satisfy the asymptotic Gilbert-Varshamov bound by using Hull dimension spectrum of linear codes and their dual codes. Carlet \cite{9} et al. characterized LCD codes and studied the structure of LCD codes and their subsets, and the results showed that almost all binary LCD codes belong to odd-like codes with odd-duals. These studies have extended the discussion on the overall structure of LCD codes. In addition, Carlet et al. \cite{10} also revealed that for $q > 3$, any $q$-ary linear code can be represented equivalently as an LCD code on $\bF_q$. In other words, in the study of LCD codes, it is enough to study only binary and ternary cases, simplifying the scope of the study.
LCD cyclic codes are also known as reversible codes. Li \cite{cyc} further analyzes the code by studying its reversibility and can improve its minimum distance lower bound.

In practical applications, LCD codes are not only used for communication error correction, but also widely used in the field of cryptography. Based on LCD codes, a new paradigm for side channel defense and resistance to fault injection attacks (\cite{11,12}), orthogonal direct and mask algorithm, is proposed. The algorithm uses the generator matrix of LCD code to encode the original ciphertext, and uses its check matrix to encode a random vector to generate mask. In other words, the algorithm takes advantage of the properties of LCD codes to divide the linear space into two complementary subspaces: one for function operations and the other for generating random masks. Because of the characteristic of the error correcting code, the algorithm can also detect whether there is an attack and produce an error during calculation. According to the provable security, if the LCD code parameter is $[n,k,d]$, the algorithm can resist the unary attack of order less than $d$. The orthogonal straight sum mask algorithm based on LCD code also has the properties of Boolean addition mask scheme, lookup table mask scheme and low entropy mask scheme, so it has high availability.

In the stage of key distribution and key recovery, Ghosh et al.\cite{16} proposed an excellent secret sharing scheme based on LCD code by using the characteristics of LCD code, which could outperform other secret sharing schemes based on other linear codes while the information rate was close to 1. In addition, the orthogonality of LCD codes makes them of great value in constructing quantum error-correcting codes (\cite{13, 14}) that can protect quantum information from noise, thus greatly advancing the development of quantum computers. Rajput et al.\cite{15} were also able to provide a solution for single node errors in distributed data storage by establishing a link between local recoverable codes and LCD codes.

\subsection{Related work: reinforcement learning}
RL is a learning mechanism whose goal is to learn how to map states to actions in order to get the maximum reward. However, RL in the initial stage faces the challenge of large state space in solving complex problems. With the rise of neural networks, the combination of RL and deep learning can better solve complex problems, thus making RL widely used \cite{17}. Vlad Mnih et al.\cite{18} first proposed the Deep Q Network (DQN) algorithm, which has pioneering significance in the field of deep reinforcement learning.

In addition to value-based methods, policy-based methods are also an important direction of reinforcement learning. The policy-based approach directly uses approximators to approximate and optimize the strategy, and finally obtains the optimal policy. This method is more suitable for continuous high dimensional action space and random policy. In 1992, Williams et al.\cite{19} proposed the REINFORCE algorithm, unifying the form of the policy gradient (PG). Subsequently, AC algorithm \cite{20} combined the value-based method and the policy-based method, and used Q value to reduce the variance of the policy, which helped to learn the policy function better. Huang et al.\cite{5} proposed a general constructor-evaluator framework based on previous studies and constructed binary linear block codes using REINFORCE algorithm and polarization codes using A2C algorithm. The data show that the performance of these learned codes is comparable to that of existing classical codes.

\subsection{Related work: Curiosity module}
Exploration and utilization are the core problems of RL. In the face of huge state space, an efficient exploration method is essential. In order to better explore, a common approach is to encourage additional exploration by means of an additional reward signal, namely an update strategy with a reward signal $r=r_e+\beta*r_i$ consisting of two parts, where $r_e$ is the external reward for environmental feedback, $r_i$ is the exploration reward, and $\beta$ is the parameter used to balance the exploration. The intrinsic reward model can be divided into two main approaches: count-based and curiosity-based. Count-based exploration measures state novelty by counting or estimating the number of states that have already been visited, encouraging the agent to visit the new state. However, for extremely large state spaces such as code space, count-based exploration is more laborious than curiosity-based exploration. The curiose-based intrinsic reward is designed to simulate environmental dynamics, which can be expressed as an error in predicting the consequences of the agent's actions, and gives a higher intrinsic reward to areas that are poorly simulated.

In part, this intrinsic reward is inspired by psychology's intrinsic motivation, just as curiosity-driven exploration may be an important way for children to grow and learn. Curiosity is essentially an internal reward, the error of the agent in predicting the consequences of its own actions in the current state. The RND (Random Network Distillation) algorithm stands out among curiosity-driven exploration methods, particularly well-suited for scenarios with vast state spaces. It leverages the discrepancy between the outputs of two neural networks to gauge the novelty of a state. Specifically, two neural networks are randomly generated, one is the target network, the parameters do not participate in the update, and the other is the prediction network, which will be constantly updated in the iteration, that is, the prediction network will fit the target network in the continuous training, and the resulting loss or the gap in the output vector can be used as an assessment of novelty. The greater the loss, the less the number of states encountered. The RND algorithm is shown in Figure 1.

\begin{center}
  \centerline
  {\includegraphics[scale=0.3]{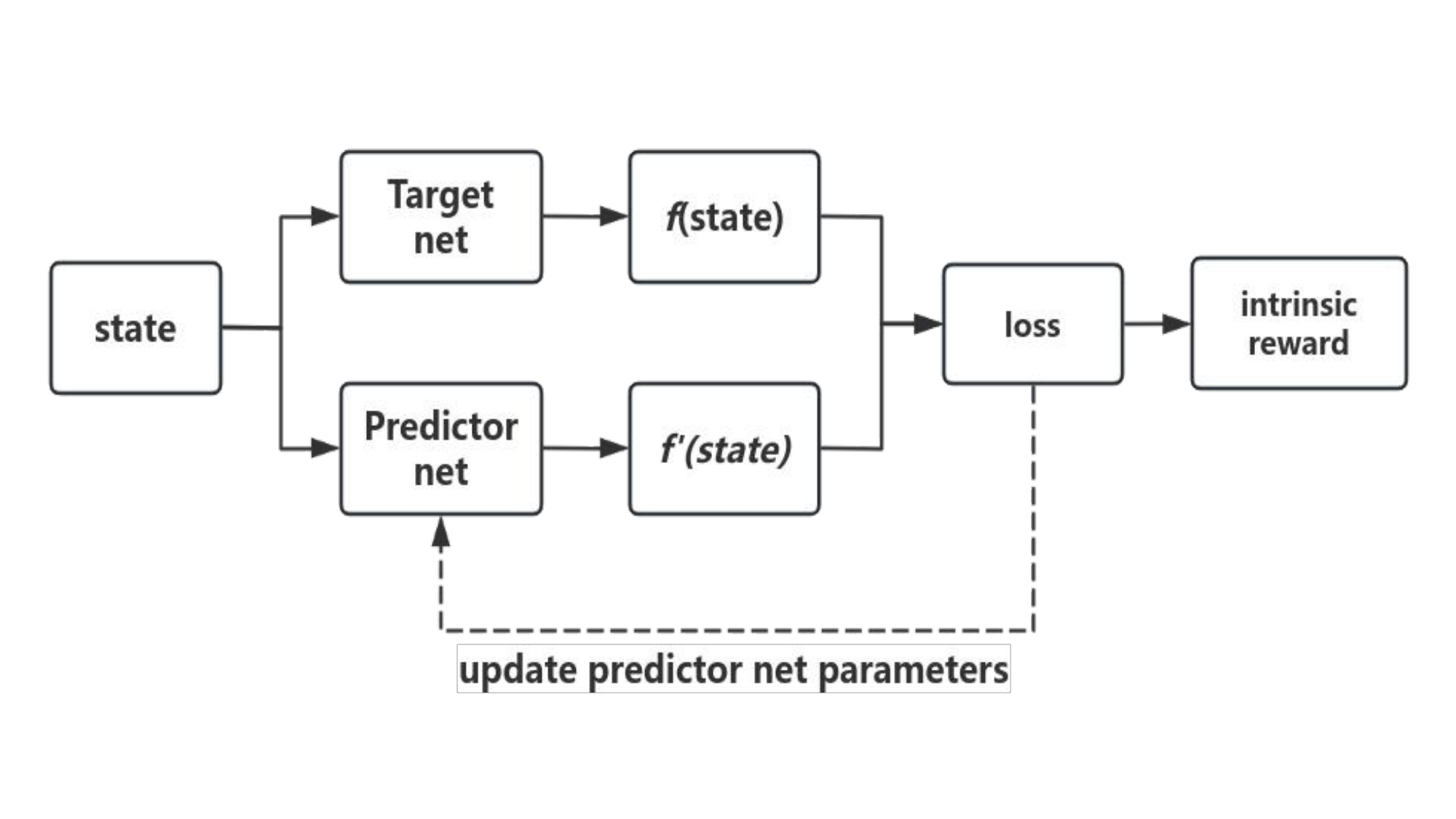}}
\centering{\small {\bf Figure 1}\ \ RND algorithm \label{rndalg}}
\end{center}

\section{LCD codes construct based on PG algorithm}
\subsection{Constructor-Evaluator}

In this paper, the constructor-evaluator framework is used to construct LCD codes. In this framework, the construction module uses RL algorithm to construct the generator matrix of LCD code. The evaluation module uses the decoder (or other performance estimation) to evaluate the performance of the LCD code, and combines with the intrinsic incentive for reprocessing to form a reward. The evaluation module feeds the reward back to the construction module, which updates itself based on the reward and continues to construct the LCD code. Repeat this process until the performance indicator converges. 
The specific frame is shown in Figure 2.

\begin{center}
  \centerline
  {\includegraphics[scale=0.3]{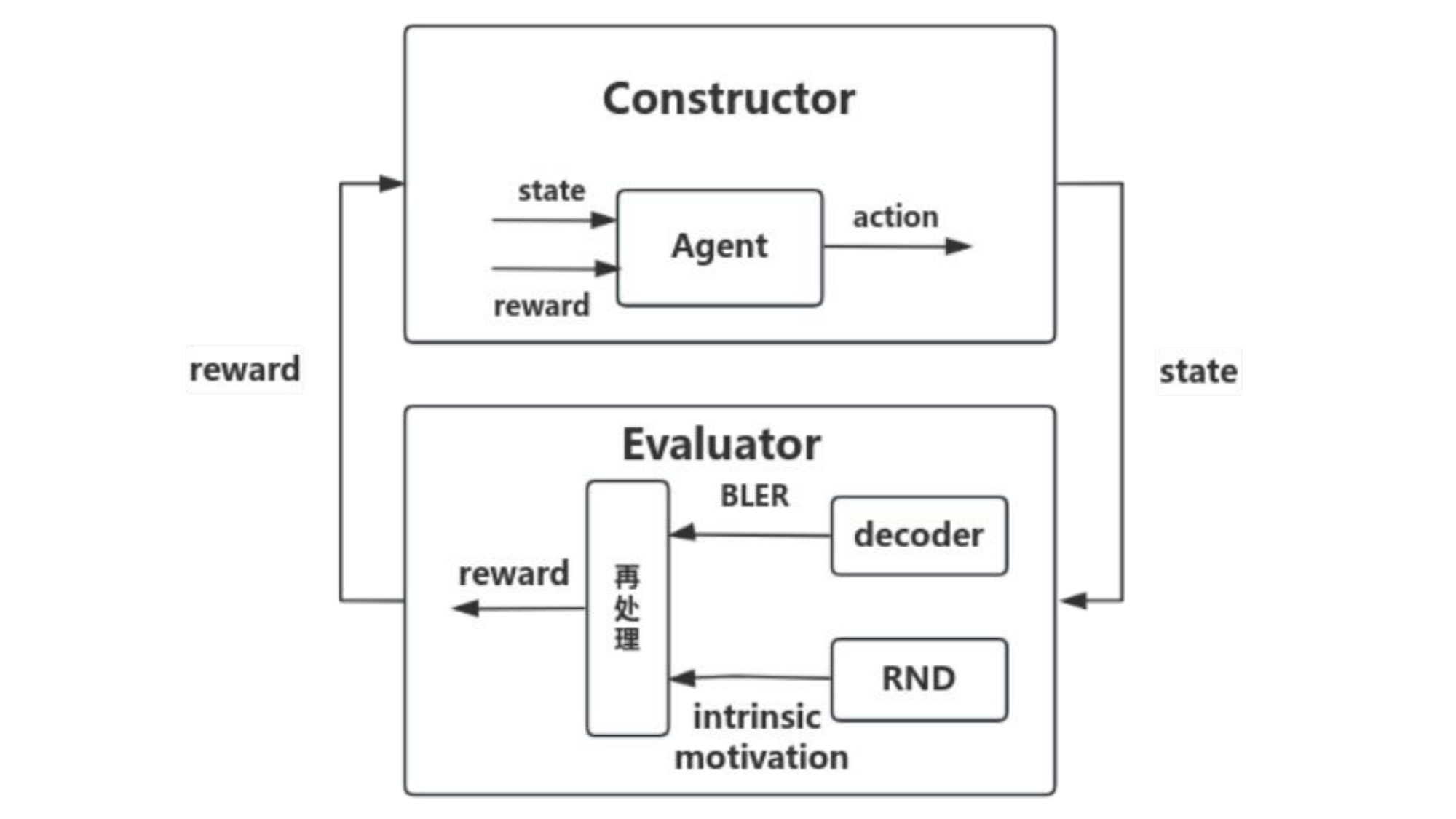}}
\centering{\small {\bf Figure 2}\ \ Constructor-Evaluator Frame \label{frame}}
\end{center}

\subsection{Constructor}
The construction process of LCD code is regarded as a decision process. This process is similar to the Markov decision process (MDP) and is defined as a quadruple ($S,A,P,r$). Where, the state $s$ represents the standard form generator matrix of the code; Action $a$ indicates the modification of the code; p stands for state transition function; $r$ represents the reward returned by the evaluation module, and the reward is positively correlated with performance. The essence of RL is to determine an optimal policy to maximize the expected return. As the construction module continuously optimizes its decision, the performance of the constructed code is improved in continuous iteration. Finally, the desired code structure can be obtained from the final state of the MDP.

The policy-based approach first needs to parameterize the policy. Therefore, the target policy $\pi_\theta$ is assumed to be random and differentiable everywhere, where $\theta$ represents the relevant parameters. Then the objective function of policy learning is defined as:

\ay
\begin{eqnarray}\label{key1}
J(\theta)=E_{s_0} (V^{\pi_\theta}(s_0))
\end{eqnarray}
\ay

Where $s_0$ represents the initial state and $V^{\pi_\theta}$ represents the state value function. Then take the derivative of the objective function to ${\theta}$, get the derivative, use the gradient ascending method to maximize the objective function, so as to get the optimal policy. After collecting certain samples, update policy is adopted by \eqref{key2}:
\ay
\begin{eqnarray}\label{key2}\nabla _{\theta} J(\theta )=E _{\pi _\theta }[\sum_{t = 0}^T(\sum_{t' = t}^T{\gamma^{(t'-t)}}r_{t'})
    {\nabla _\theta}\log {\pi _\theta }(a_t | s_t) ]
\end{eqnarray}
\ay

Where $T$ is the maximum number of steps interacting with the environment, and $\gamma$ is the discount factor.

To solve MDP, the REINFORCE algorithm was used, which was a classic algorithm of the policy gradient method in RL. The REINFORCE algorithm theoretically ensured local optimization and used the Monte Carlo method to sample trajectories to estimate the action value. The advantage of this method was that an unbiased gradient could be obtained. In addition, using the powerful approximation ability of neural network, the policy function is represented by neural network. Then the neural network is updated by gradient descent to promote the optimization of the policy function to improve the performance.

\subsection{Evaluator}

The role of the evaluator includes evaluating the performance of the constructed code, generating intrinsic incentives and determining whether it is an LCD code, and converting the performance indicators into appropriate rewards. The reward from the evaluator is the key communication link between human intent and algorithm execution, communicating the desired outcome to the AI and guiding the direction in which the AI learns. By setting the right rewards and punishments, AI can effectively construct LCD codes while avoiding the construction of substandard codes.

Specifically, the evaluator will first determine whether the constructed code is LCD code, if it is LCD code, it will evaluate the block error rate (BLER) of the constructed LCD code at the fixed signal-to-noise ratio (SNR) point through the hierarchical ordered decoding (OSD) device, and then evaluate the performance of the LCD code. Certainly, other performance metrics for LCD codes are also viable for evaluation. However, at the beginning, AI does not understand coding theory and does not understand LCD codes, and may construct non-LCD codes, in order to avoid this situation, AI will be given corresponding penalties when constructing non-LCD codes. Then the generator matrix of the constructed code is input into the RND algorithm module to evaluate the novelty of the state, and output the novelty as an intrinsic incentive. The RND module updates itself based on this state. Finally, the evaluator combines the external reward and the internal incentive, and feeds back to the construction module in the form of reward. This reward mechanism helps guide the AI to learn the process of constructing LCD codes, so that it gradually understands and follows the coding theory, while maintaining the novelty of the construction code.

\section{Some examples of binary LCD codes}

Suppose that the generator matrix of a binary LCD code $\MC$ with length $n$ and dimension $k$ is $G$. Meanwhile $G=[I_k,P]$ is a standard generator matrix, where $I_k$ is an identity matrix of size $k\times k$, and $P$ is an ordinary matrix of size $k\times(n-k)$. Let $\bF_q$ be a finite field of order $q$, where $q$ is an odd prime number.
An $[n,k]_q$ linear code $\MC$ in $\bF_q$ is a subspace of $\bF^n_q$. We call an  $[n,k,d]_q$ code $\MC$  optimal if there is no $[n,k,d+1]_q$ code.   An $[n,k,d]_q$ code is called almost optimal if there exists an  optimal $[n,k,d+1]_q$ code  \cite{21}. The dual code of a code $C$ of length $n$ in $\bF_q$ is defined as:
$$\mathcal C^{\bot} = \{ {\bf x}\in \Bbb F_q^{n} \mid \langle {\bf x},{\bf y}\rangle=0~  \forall~  {\bf y}\in \mathcal C\},$$
where ${\bf x}=(x_0, \ldots, x_{n-1})$, $ {\bf y}= (y_0, \ldots,y_{n-1})$, and $ \langle {\bf x},{\bf y}\rangle=\sum_{i=0}^{n-1}x_iy_i.$ A linear code $\mathcal C$ over $\Bbb F_q$ is called a {\em  Linear Complementary Dual (LCD)
code } if $\mathcal C \cap  \mathcal C^{\bot} = \{0\}$.

The following lemma is very important for  LCD codes. 

\begin{lemma}[see \cite {22}] \label{lem1} 
Let $G$ denote the generator matrix of a code $\mathcal C$. 
Then  $\mathcal C$ is LCD if and only if 
$GG^T$ is nonsingular, where $G^T$ is the transpose of $G$.
\end{lemma}

We're going to restrict AI to constructing codes that satisfy Lemma \ref{lem1}. Mainly, soft constraints are enforced through the use of rewards.

\subsection{Binary LCD codes}
We denote the finite field of order 2 by $\bF_2 = \{0,1\}$,  an $[n, k, d]$ LCD code over $\bF_q$ by $\mathcal C$, and its generator matrix by $[I_k | \mathcal P_q^{n,k}]$.
We use a fully connected neural network with one input layer, two hidden layers and one output layer. Details of each layer of the network are as follows:

\begin{itemize}
    \item[$\bullet$] The input to the neural network is the generator matrix of binary LCD codes. Since the size of the generator matrix is $k \times n$, the number of input nodes of the neural network is $kn$. Set the initial status to $s_0=[I_k|0]$.
    
    \item[$\bullet$] The output node number of the neural network is $k(n-k)$. Action $a$ is a real matrix of size $k \times (n-k)$ sampled from the policy $\pi_\theta$. In order to construct a binary LCD code, sigmoid activation function is used to normalize the size of the action to the range of $0-1$. If the value is greater than 0.5, element 1 in the binary field is selected; otherwise, element 0 in the binary field is selected.
    
    \item[$\bullet$] The number of hidden layers is $2k(n-k)$, and there are two hidden layers. Two hidden layers with appropriate activation functions can represent any decision boundary with any precision and can fit any smooth map with any precision.
    
    \item[$\bullet$] The reward is set at $1-\lambda*BLER+\beta*r_i$. Note that here $\lambda$ is a constant, BLER is the block error rate calculated by OSD decoder, and $r_i$ is the internal incentive generated by RND module. However, AI does not understand coding theory, nor does it understand LCD codes, and may construct non-LCD codes during the construction process. In this case, very small rewards are used to guide the agent. If the obtained code is non-LCD, which means that $GG^T$ is a singular matrix, then the reward will be equal to $-1+\beta*r_i$.
    
\end{itemize}
The PG algorithm for LCD code construction is also described in {\bf Algorithm \ref{alg1}}.
\begin{algorithm}
    \caption{Policy gradient based LCD codes design}\label{algorithm}
    \label{alg1}
    \begin{algorithmic}
    \WHILE{1}
        \STATE Set initial state $s_0 = [I_k|0]$, $\sigma ^2 =0.1$
        \FOR{$step = {1,\cdots ,maxstep}$}
            \STATE $\pi _\theta  \leftarrow NeuralNet(\theta ) $
            \FOR{$i={1,\cdots ,k}$}
                \FOR{$j={1,\cdots ,n-k}$}
                    \STATE $a_{i,j}\sim {N ({\pi _\theta}_{i,j}, \sigma ^2)}$
                    \STATE $P_{i,j}=(a_{i,j}\geqslant 0.5)?1:0$
                \ENDFOR
            \ENDFOR
            \STATE $s_1\leftarrow [I_k|P]$
            \STATE $r\leftarrow  Evaluator(s_1)$
            \STATE $\pi_\theta \leftarrow f_N(a|\mu,\sigma^2)$
        \ENDFOR
        \STATE $\theta \leftarrow agent.learn$
    \ENDWHILE
    \end{algorithmic}
\end{algorithm}

In this paper, the Tianshou RL framework \cite{23} is adopted, and the Adam optimizer and the update policy based on small-batch random gradient descent are used in the algorithm. The performance comparison between the constructed LCD code and the code constructed based on coding theory and the common linear code constructed based on RL is shown in Figure 3.

\begin{center}
  \centerline
  {\includegraphics[scale=0.5]{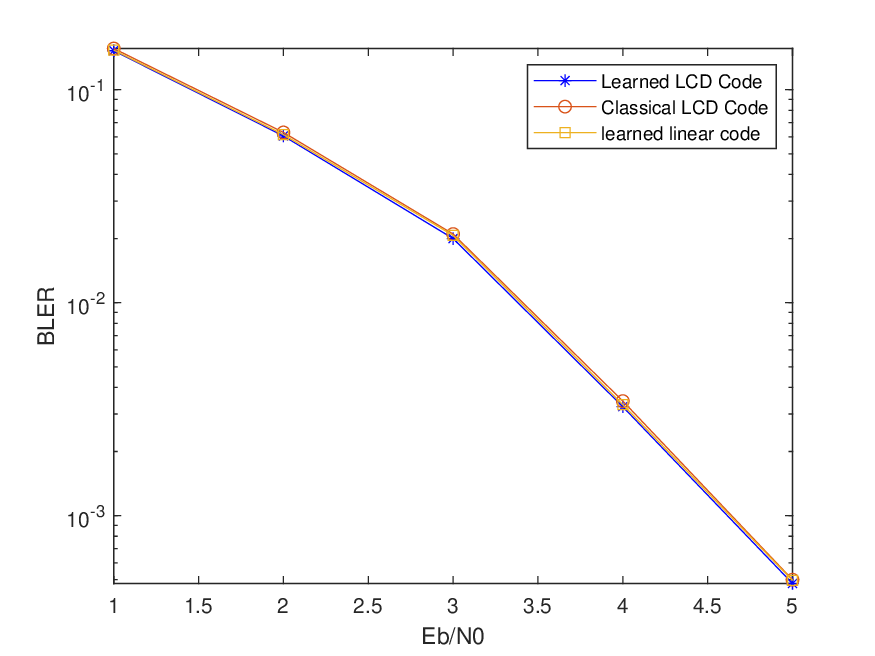}}
\centering{\small {\bf Figure 3}\ \ BLER comparison \label{bler}}
\end{center}

Figure 3 shows the comparison of BLER calculated by OSD decoder for LCD codes constructed based on RL, almost optimal LCD codes constructed based on coding theory and ordinary linear codes constructed based on RL under AWGN channel condition when the code length is $n=24$ and the dimension is $k=12$. Among them, the  almost optimal LCD codes constructed based on coding theory are constructed on the basis of $[23,11]$ LCD code \cite{24}. The BLER of LCD codes constructed based on RL is slightly lower than that of almost optimal LCD codes constructed based on coding theory, which indicates that the performance of LCD codes constructed based on AI technology can not only reach, but even surpass the performance of LCD codes constructed based on classical construction algorithms based on coding theory under some channel conditions. In addition, compared with the method based on coding theory, the construction method based on RL does not need to rely on researchers' manual calculation and derivation, can realize automatic coding, and does not need to pay attention to the constraint between code length and dimension, so it is very flexible.
The BLER of LCD code constructed based on RL in AWGN channel is lower than that of common linear code constructed by RL, that is, when using RL to construct linear code, adding structure to the code can further improve the performance of the constructed code. Compared with ordinary linear codes, LCD codes constructed by RL can be used not only in data error detection and error correction, but also in the construction of codes with other structures. For example, in coding theory, $[n+1,k+1]$ LCD codes, quantum error correction codes and linear complementary pair of codes (LCP) can be constructed on the basis of $[n,k]$ LCD codes. The usability of the constructed code is improved.

In coding theory, in addition to length $n$ and dimension $k$, minimum distance $d$ is also an important parameter of linear code, so $[n,k,d]$ is often used to represent its parameters. There is a constraint relationship between the three parameters. However, when $n$ and $k$ are fixed, the larger $d$ is, the better the performance of the code. Theoretically, the error correction code can check $d-1$ errors and correct $\lfloor \frac{d-1}{2} \rfloor$ errors.

After experiments, using the minimum distance $d$ as a reward can also construct an LCD code, and the minimum distance of the constructed LCD code will be as large as possible, and sometimes even reach the minimum distance limit.

\begin{center}
  \centerline
  {\includegraphics[scale=0.5]{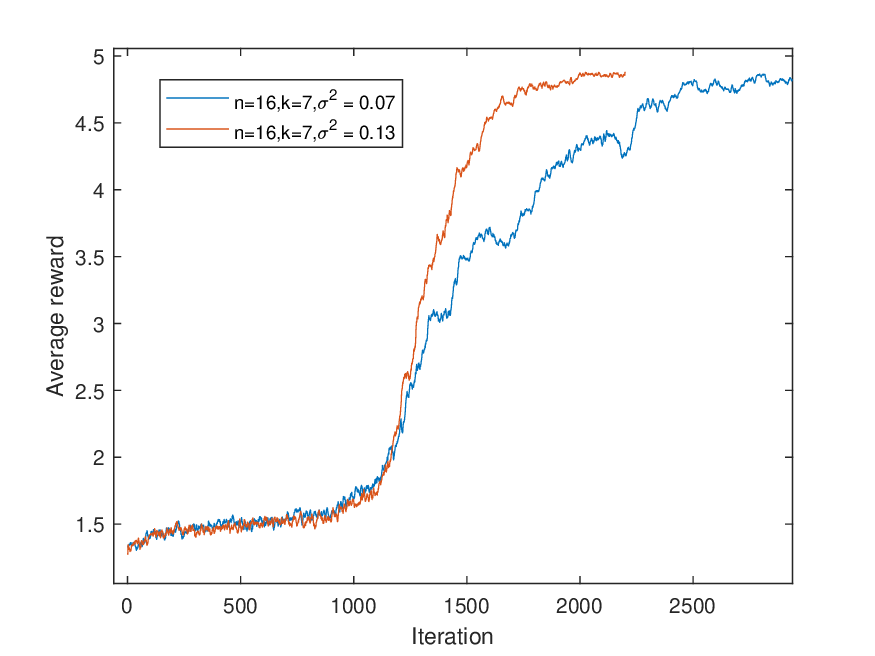}}
\centering{\small \ {\bf Figure 4}\ Comparison of different variances \label{variances}}
\end{center}

In Figure 4, we compare the influence of $\sigma^2$ value on the experiment. The figure shows that reducing the value of $\sigma^2$ will slow down the learning process; On the contrary, appropriately increasing the value of $\sigma^2$ will have a certain effect on the whole process, but too large $\sigma^2$ will have the opposite effect. This is because $\sigma^2$ controls the degree of exploration of RL in the construction process, and appropriate setting of parameters can accelerate the learning process.

When the minimum distance is rewarded, the construction method based on RL can construct the LCD code with the minimum distance as large as possible, and even reach the minimum distance boundary. At present, the minimum distance bound of some LCD codes is not clear, only the range of existence of the bound is clear (\cite{2,24}). Through the proposed method, the existence of some almost optimal LCD codes is found to further shorten the range of minimum distance bounds. The specific results are shown in Table 1 and several examples below. Table 1 records the minimum distances and upper bounds of some constructed binary LCD codes:

\begin{center}
{\small
    \begin{center}
    \centerline{\small {\bf Table 1}\ \  Minimum distances of learned binary LCD codes }\vskip 1mm
    \label{binary}\tabcolsep 15pt

    {\small
    \begin{tabular}{cccc}
        \toprule
        \textbf{n}& \textbf{k} & \textbf{bound} & \textbf{learned distance} \\
        \midrule
        19 & 13 & 3 & 3 \\ 
	    23 & 10 & 7 & 6 \\  
	    24 & 11 & 7 & 6 \\  
	    29 & 11 & 8-9 & 8 \\ 
	    33 & 12 & 10 & 9 \\ 
	    37 & 9 & 12-14 & 13 \\ 
	    39 & 9 & 12-16 & 14 \\ 
	    40 & 9 & 12-16 & 13 \\ 
	    40 & 10 & 12-16 & 13 \\ 
        \bottomrule
    \end{tabular}}
    \end{center}}
    \end{center}

\begin{example} 
\rm Let $\mathcal C$ be an $[37,9]$ binary LCD code. As we know,  optimal minimum distance of $\mathcal C$ should be  between 12 and 14. We find the existence of an LCD code with minimum distance of 13.
The $[37,9,13]$ binary LCD code $\mathcal C$ has a generator matrix $(I_9|\mathcal P_2^{37,9})$

\setcounter{MaxMatrixCols}{40}

\setlength{\arraycolsep}{3pt}
\begin{center}$
\mathcal P_2^{37,9}=\left [\begin{matrix}
1& 1& 1& 1& 1& 0& 0& 0& 0& 0& 0& 1& 1& 0& 0& 0& 1& 0& 1& 1& 1& 0& 0& 0& 1& 0& 0& 1\\
0& 0& 1& 0& 1& 0& 0& 0& 0& 1& 0& 0& 1& 0& 1& 1& 1& 1& 1& 0& 0& 1& 1& 0& 1& 0& 0& 1\\
0& 1& 1& 0& 1& 1& 0& 1& 0& 1& 0& 1& 0& 0& 0& 1& 1& 0& 0& 0& 1& 1& 0& 1& 0& 1& 0& 1\\
1& 0& 1& 0& 1& 1& 0& 1& 0& 0& 0& 1& 1& 1& 1& 1& 0& 1& 1& 0& 1& 0& 0& 0& 0& 1& 1& 1\\
1& 0& 0& 0& 1& 1& 1& 1& 1& 1& 1& 0& 1& 0& 0& 0& 0& 1& 1& 1& 1& 1& 1& 1& 0& 1& 0& 0\\
0& 0& 0& 1& 0& 1& 0& 1& 1& 1& 0& 0& 1& 0& 1& 0& 0& 0& 0& 1& 1& 1& 1& 0& 0& 0& 1& 1\\
1& 1& 0& 1& 1& 1& 1& 0& 0& 1& 1& 1& 0& 1& 0& 1& 1& 1& 0& 0& 0& 0& 1& 0& 0& 0& 1& 1\\
0& 0& 1& 1& 0& 0& 1& 0& 1& 0& 0& 1& 0& 1& 1& 1& 1& 1& 1& 0& 1& 1& 0& 1& 1& 1& 1& 0\\
1& 1& 1& 1& 1& 0& 0& 1& 1& 0& 1& 0& 0& 1& 0& 0& 1& 1& 1& 0& 0& 1& 1& 0& 1& 1& 1& 1\\
\end{matrix} \right].$
\end{center}

\end{example}

\begin{example} 
\rm Let $\mathcal C$ be an $[39,9]$ binary LCD code. As we know,  optimal minimum distance of $\mathcal C$ should be  between 12 and 16. We find the existence of an LCD code with minimum distance of 14.
The $[39,9,14]$ binary LCD code $\mathcal C$ has a generator matrix $(I_9|\mathcal P_2^{39,9})$ 
\setcounter{MaxMatrixCols}{40}

\setlength{\arraycolsep}{3pt}
\begin{center}$
\mathcal P_2^{39,9}=\left [\begin{matrix}
1& 1& 1& 1& 1& 0& 1& 0& 0& 1& 1& 0& 1& 1& 1& 0& 0& 1& 1& 1& 1& 0& 1& 0& 1& 0& 0& 0& 1& 1\\
1& 1& 1& 0& 0& 1& 0& 1& 0& 1& 1& 1& 1& 1& 1& 0& 1& 0& 1& 0& 1& 0& 1& 0& 1& 1& 1& 0& 0& 0\\
0& 1& 0& 0& 1& 1& 1& 0& 1& 1& 0& 1& 0& 1& 1& 1& 0& 1& 1& 1& 1& 0& 1& 1& 0& 0& 1& 1& 0& 0\\
1& 1& 1& 0& 0& 0& 1& 0& 1& 1& 1& 0& 0& 0& 0& 1& 1& 1& 0& 0& 0& 1& 1& 1& 1& 0& 1& 0& 1& 0\\
0& 1& 0& 1& 1& 0& 1& 1& 1& 1& 1& 0& 0& 0& 0& 1& 1& 0& 0& 0& 1& 1& 0& 0& 1& 0& 0& 0& 0& 1\\
0& 1& 1& 1& 1& 0& 0& 0& 1& 0& 0& 0& 1& 1& 1& 0& 0& 0& 0& 1& 0& 0& 1& 1& 0& 1& 1& 1& 1& 1\\
1& 1& 0& 0& 1& 1& 0& 1& 0& 1& 1& 0& 1& 1& 0& 0& 1& 0& 0& 1& 0& 1& 1& 1& 0& 0& 1& 0& 0& 1\\
0& 0& 0& 0& 0& 1& 0& 1& 0& 1& 1& 0& 1& 0& 1& 0& 0& 1& 1& 1& 1& 1& 0& 0& 0& 1& 0& 1& 1& 1\\
1& 1& 0& 0& 0& 1& 0& 1& 0& 1& 0& 1& 0& 0& 0& 1& 0& 1& 0& 0& 0& 0& 1& 1& 0& 1& 0& 1& 1& 1\\
\end{matrix} \right].$
\end{center}

\end{example}

\begin{example} 
\rm Let $\mathcal C$ be an $[40,9]$ binary LCD code. As we know,  optimal minimum distance of $\mathcal C$ should be  between 12 and 16. We find the existence of an LCD code with minimum distance of 13.
The $[40,9,13]$ binary LCD code $\mathcal C$ has a generator matrix $(I_9|\mathcal P_2^{40,9})$ 
\setcounter{MaxMatrixCols}{40}

\setlength{\arraycolsep}{3pt}
\begin{center}$
\mathcal P_2^{40,9}=\left [\begin{matrix}
 1& 1& 1& 1& 0& 0& 0& 0& 1& 1& 1& 0& 0& 0& 1& 1& 1& 1& 1& 0& 1& 0& 0& 0& 0& 1& 1& 0& 1& 1& 1\\
 1& 0& 0& 1& 0& 1& 1& 0& 0& 1& 1& 1& 1& 1& 1& 0& 0& 0& 0& 0& 0& 1& 1& 1& 1& 0& 1& 1& 1& 1& 0\\
 1& 1& 1& 0& 0& 1& 1& 1& 0& 0& 1& 1& 1& 1& 0& 1& 0& 0& 0& 1& 0& 0& 1& 0& 0& 1& 0& 0& 0& 0& 0\\
 1& 0& 1& 0& 0& 0& 1& 1& 0& 0& 0& 0& 1& 0& 0& 0& 1& 1& 1& 1& 0& 0& 1& 1& 1& 1& 0& 1& 1& 1& 0\\
 0& 1& 1& 0& 0& 1& 0& 0& 1& 1& 1& 0& 1& 1& 0& 0& 0& 0& 1& 0& 0& 1& 1& 1& 0& 1& 0& 0& 0& 1& 1\\
 0& 0& 0& 1& 0& 1& 0& 1& 0& 0& 0& 1& 0& 0& 1& 1& 1& 0& 1& 1& 1& 0& 0& 1& 1& 0& 0& 1& 1& 0& 1\\
 1& 0& 1& 1& 1& 0& 1& 0& 1& 0& 1& 1& 1& 0& 1& 1& 0& 1& 1& 1& 0& 1& 1& 0& 1& 1& 1& 0& 1& 0& 1\\
 1& 1& 0& 0& 1& 1& 0& 1& 1& 1& 1& 1& 1& 0& 0& 1& 0& 0& 1& 0& 0& 0& 0& 0& 1& 0& 0& 1& 1& 0& 0\\
 0& 0& 1& 0& 1& 1& 1& 0& 1& 1& 0& 0& 1& 0& 1& 0& 1& 1& 1& 0& 1& 1& 0& 1& 0& 0& 0& 1& 1& 1& 1\\
\end{matrix} \right].$
\end{center}

\end{example}

\subsection{Some examples of ternary LCD codes}
For LCD codes, the examination of both binary and ternary systems is found to be adequately enlightening, thereby motivating our endeavor to explore the realm of ternary LCD code construction.
In the binary case, sigmoid activation functions are commonly used for binary classification. 
Therefore, we assumed that by establishing an appropriate mapping between actions and states, we could construct ternary LCD codes in a similar manner.
And we denote the finite field of order 3 by $\bF_3 = \{0,1,2\}$.

The algorithm for constructing ternary LCD codes is similar to that for constructing binary LCD codes, with one key difference.
Specifically, we set the activation function of the output layer to softplus and apply a linear transformation to amplify its output values.
This approach is necessary because we will be using modulo 3 to generate three elements in the finite field.

\begin{algorithm}
    \caption{PG based ternary LCD codes design}\label{algorithm2}
    \begin{algorithmic}
    \WHILE{1}
        \STATE Set initial state $s_0 = [I_k|0]$, $\sigma ^2 =0.3$
        \FOR{$step = {1,\cdots ,maxstep}$}
            \STATE $\mu  \leftarrow NeuralNet(\theta ) $
            \FOR{$i={1,\cdots ,k}$}
                \FOR{$j={1,\cdots ,n-k}$}
                    \STATE $a_{i,j}\sim {N (\mu_{i,j}, \sigma ^2)}$
                    \STATE $\mathcal P_{i,j}=\lceil a_{i,j} \rceil \bmod 3$
                \ENDFOR
            \ENDFOR
            \STATE $s_1\leftarrow [I_k|\mathcal P]$
            \STATE $r\leftarrow  Evaluator(s_1)$
			\STATE $\pi_\theta \leftarrow f_N(a|\mu,\sigma^2)$
        \ENDFOR
        \STATE $\theta \leftarrow Update ~by ~Equation~ (2)$
    \ENDWHILE
    \end{algorithmic}
\end{algorithm}

Similarly, the minimum range bounds of some ternary LCD code are uncertain. Through {\bf Algorithm \ref{algorithm2}}, some ternary LCD codes are constructed to shorten the range of the current bounds, and even reach the bounds. Detailed data are shown in Table 2.

\begin{center}
{\small
    \begin{center}
    \centerline{\small {\bf Table 2}\ \ Minimum distances of learned ternary LCD codes }\vskip 1mm
    \label{ternary}\tabcolsep 15pt

    {\small
    \begin{tabular}{cccc}
        \toprule
        \textbf{n}& \textbf{k} & \textbf{bound} & \textbf{learned distance} \\
        \midrule
        20 & 5 & 11 & 10 \\ 
	    22 & 4 & 12 & 11 \\  
	    22 & 4 & 12-13 & 13 \\  
	    22 & 5 & 11-12 & 12 \\ 
	    23 & 4 & 13-14 & 14 \\ 
	    23 & 5 & 11-13 & 12 \\  
	    24 & 6 & 11-13 & 12 \\ 
	    25 & 4 & 15-16 & 15 \\
	    25 & 5 & 13-15 & 14 \\ 
        \bottomrule
    \end{tabular}}
    \end{center}}
    \end{center}

\begin{example} 
\rm Let $\mathcal C$ be an $[22,4]$ ternary LCD code. As we know, optimal minimum distance of $\mathcal C$ should be between 12 and 13. We find the existence of an LCD code with minimum distance of 13.
The $[22,4,13]$ binary LCD code $\mathcal C$ has a generator matrix $(I_4|\mathcal P_3^{22,4})$ 
\begin{center}$
\mathcal P_3^{22,4}=\left [\begin{matrix}
0& 0& 1& 2& 1& 1& 1& 1& 0& 1& 1& 1& 2& 1& 1& 2& 0& 2\\
1& 1& 2& 1& 1& 1& 0& 1& 0& 0& 1& 0& 1& 2& 0& 0& 1& 1\\
1& 0& 2& 1& 1& 2& 1& 0& 1& 1& 1& 2& 0& 0& 0& 1& 2& 1\\
1& 1& 1& 1& 0& 0& 1& 0& 1& 1& 1& 1& 1& 1& 2& 0& 0& 0\\
\end{matrix} \right].$
\end{center}

\end{example}

\begin{example} 
\rm Let $\mathcal C$ be an $[23,5]$ ternary LCD code. As we know, optimal minimum distance of $\mathcal C$ should be between 11 and 12. We find the existence of an LCD code with minimum distance of 12.
The $[23,5,12]$ binary LCD code $\mathcal C$ has a generator matrix $(I_5|\mathcal P_3^{23,5})$ 

\setlength{\arraycolsep}{3pt}
\begin{center}
$\mathcal P_3^{23,5}=\left [\begin{matrix}
 0& 1& 2& 1& 1& 0& 1& 0& 0& 1& 1& 2& 1& 1& 1& 2& 1& 2\\
 1& 2& 0& 1& 1& 1& 1& 0& 1& 2& 1& 0& 0& 2& 1& 1& 2& 1\\
 1& 1& 0& 0& 1& 1& 0& 1& 1& 1& 1& 1& 0& 1& 2& 0& 0& 1\\
 1& 1& 0& 1& 0& 1& 1& 1& 0& 0& 2& 1& 1& 1& 1& 1& 1& 1\\
 2& 0& 1& 1& 2& 1& 2& 1& 1& 1& 1& 0& 1& 1& 1& 2& 1& 1\\
\end{matrix} \right].$
\end{center}

\end{example}

\begin{example} 
\rm Let $\mathcal C$ be an $[24,6]$ ternary LCD code. As we know, optimal minimum distance of $\mathcal C$ should be between 11 and 13. We find the existence of an LCD code with minimum distance of 12.
The $[24,6,12]$ binary LCD code $\mathcal C$ has a generator matrix $(I_4|\mathcal P_3^{24,6})$ 

\setlength{\arraycolsep}{3pt}
\begin{center}
$\mathcal P_3^{24,6}=\left [\begin{matrix}
0& 1& 2& 2& 0& 2& 1& 1& 1& 1& 1& 1& 2& 0& 1& 2& 1& 0\\
1& 2& 1& 0& 1& 0& 1& 1& 1& 2& 1& 1& 0& 1& 2& 0& 2& 1\\
0& 1& 0& 0& 2& 1& 0& 1& 2& 2& 0& 1& 0& 1& 0& 1& 1& 2\\
2& 0& 1& 2& 0& 1& 2& 1& 1& 1& 1& 2& 1& 1& 1& 0& 2& 1\\
2& 1& 1& 0& 1& 1& 0& 2& 1& 1& 1& 1& 1& 2& 2& 1& 1& 1\\
1& 1& 1& 1& 0& 0& 1& 2& 2& 1& 1& 1& 2& 1& 1& 1& 0& 1\\
\end{matrix} \right].$
\end{center}

\end{example}

\subsection{RND module}

This section verifies that RND module, as an exploration method of internal stimulation, can add curiosity to the agent and drive it to explore in the environment. In the experiment, only internal incentives were used as reward signals, and Figure 5 shows the changes of internal incentives with continuous exploration. 

\begin{center}
  \centerline
  {\includegraphics[scale=0.5]{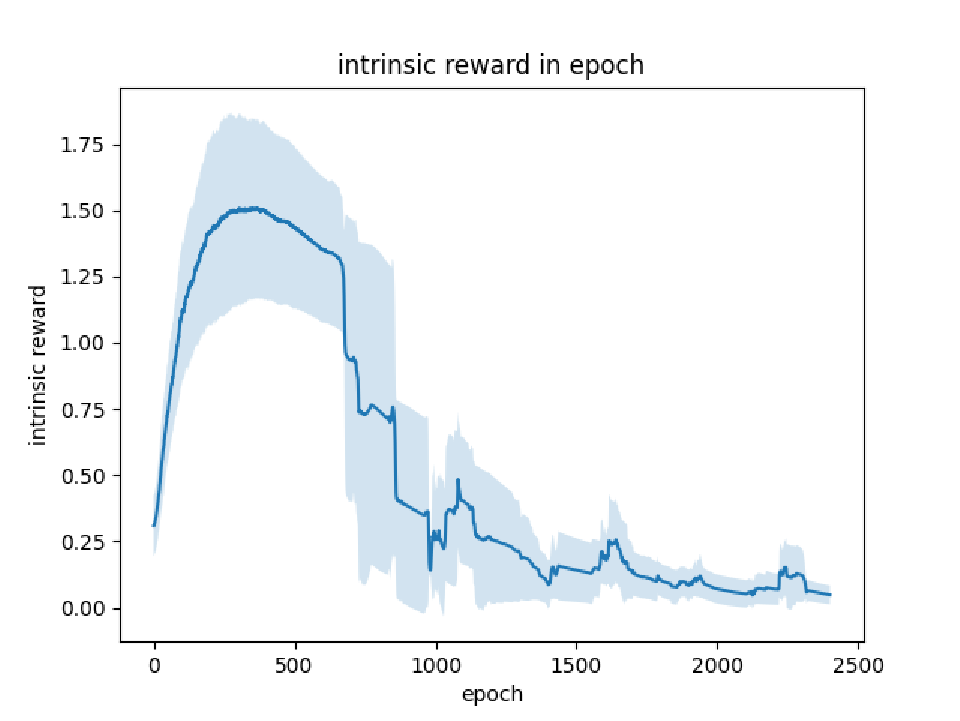}}
\centering{\small \ {\bf Figure 5}\ Intrinsic excitation variation \label{intrinsic}}
\end{center}

The first half of the curve describes that by making the agent curious about the unknown in the environment, the RND module encourages the agent to actively choose those actions that can maximize the prediction error, showing strong exploration behavior. This curiosity-driven approach to exploration makes agents more likely to discover new areas, new states, and more responsive to unknown environments. The second half of the curve shows that the agent can actively explore new behavior strategies even when facing a state similar to the known state, which helps to jump out of the local optimum.

In order to verify the influence of RND module on the construction process, we compared the construction effects of adding RND module and not adding RND module in the construction of $[20,8]$ LCD codes through ablation experiments. The contrast reward changes are shown in Figure 6.

\begin{center}
  \centerline
  {\includegraphics[scale=0.5]{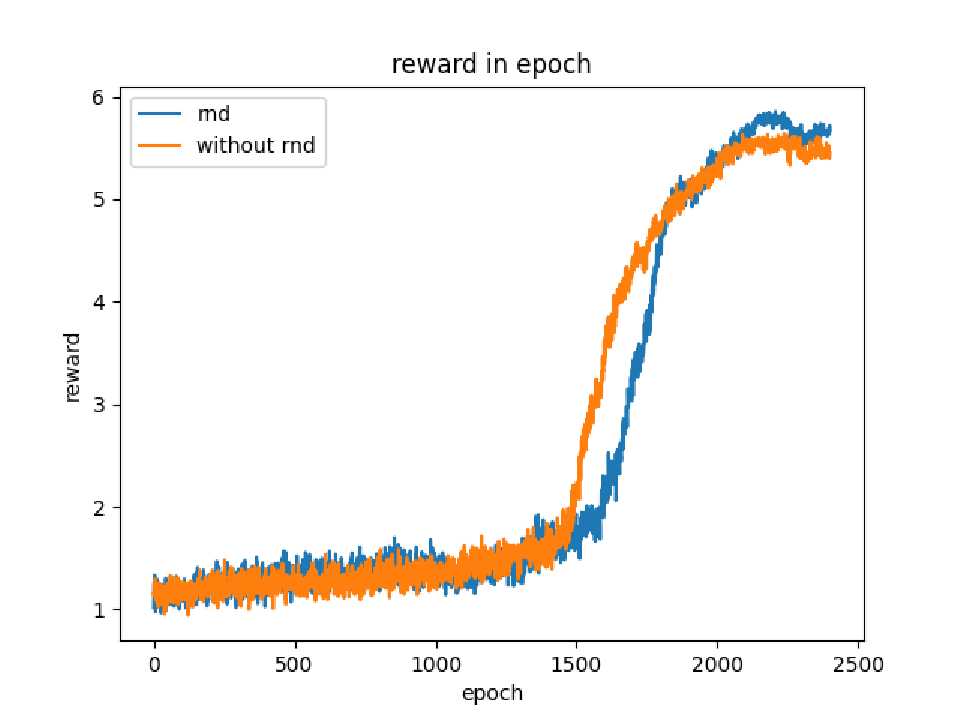}}
\centering{\small \ {\bf Figure 6}\ Performance comparison between with and without RND \label{RND_compare}}
\end{center}

The experimental results show that the number of convergent iterations with RND module is similar to that without RND module, indicating that RND module does not significantly increase the number of iteration rounds. However, with the addition of the RND module, the model achieved a higher average reward in the same number of iterations. By calculating the average reward of the last 100 rounds, it is found that the reward with RND module is 3.23\% higher than that without RND module. This result shows that the RND module improves the performance of the model during training by guiding the agent to more efficient exploration. More importantly, the introduction of the RND module seems to help avoid falling into local optimal solutions.

\section{Conclusions}

In this paper, by integrating coding theory into the reward setting of reinforcement learning, we successfully add structure to the codes constructed by reinforcement learning, and construct binary and ternary LCD codes. By setting appropriate rewards, AI can construct code with a specified structure, further improving the performance and usability of the code; The multi-component LCD code is successfully constructed by reasonably setting the mapping of action to state, which means that the AI-based method is not limited to the binary field. Through the integration of the RND module within the evaluation mechanism, we have successfully augmented the model's exploratory capacity, leading to a subsequent enhancement in model performance.

Furthermore, conventional coding theory is based on optimizing parameters for code creation, whereas AI-driven coding relies on refining algorithms to enhance code performance. This entails employing advanced RL algorithms to further optimize code structure, and the progression of RL, along with its advanced ideas, also contributes to the advancement of ECC.

\section*{Conflict of Interest}
The authors declare no conflict of interest.




\section*{Acknowledgements}
{\rm This work was supported by the National Natural Science Foundation of China (Nos. 62372247, 12101326, 62206133)  and  the Open Project of Guangxi Provincial Key Laboratory (No. MIMS22-01).}




\setcounter{equation}{0}
\renewcommand\theequation{A.\arabic{equation}}

\end{document}